\documentclass{IEEEcsmag}

\usepackage[colorlinks,urlcolor=blue,linkcolor=blue,citecolor=blue]{hyperref}

\usepackage{upmath}

\jvol{XX}
\jnum{XX}
\paper{XX}
\jmonth{XX}
\jname{Computer}
\pubyear{2020}

\begin{document}


\title{Cyberphysical Security Through Resiliency: A Systems-centric Approach}

\author{Cody Fleming}
\affil{Iowa State University}

\author{Carl Elks}
\affil{Virginia Commonwealth University}

\author{Georgios Bakirtzis}
\affil{University of Virginia}

\author{Stephen C. Adams}
\affil{University of Virginia}

\author{Bryan Carter}
\affil{University of Virginia}


\author{Peter A. Beling}
\affil{University of Virginia}

\author{Barry Horowitz}
\affil{University of Virginia}

\markboth{Cyber-Physical Security Through Resiliency: A Systems-centric Approach}{Cyber-Physical Security Through Resiliency: A Systems-centric Approach}

\begin{abstract}
Cyber-physical systems (CPS) are often defended in the same manner as information technology (IT) systems -- by using perimeter security. Multiple factors make such defenses insufficient for CPS. Resiliency shows potential in overcoming these shortfalls. Techniques for achieving resilience exist; however, methods and theory for evaluating resilience in CPS are lacking. We argue that such methods and theory should assist stakeholders in deciding where and how to apply design patterns for resilience. Such a problem potentially involves tradeoffs between different objectives and criteria, and such decisions need to be driven by traceable, defensible, repeatable engineering evidence. Multi-criteria resiliency problems require a system-oriented approach that evaluates systems in the presence of threats as well as potential design solutions once vulnerabilities have been identified. We present a systems-oriented view of cyber-physical security, termed {\em Mission Aware}, that is based on a holistic understanding of mission goals, system dynamics, and risk.
\end{abstract}

\maketitle

\chapterinitial{
Cyber-physical systems (CPS)}~\cite{serpanos:2018} are increasingly the subjects of cyber-attacks and threats. From microelectronics chips to operating systems, data networks, and wireless network protocols—threats and exploits proliferate at increasingly high rates due to adversaries, from nation-state actors to hackers~\cite{cardenas:2008,bakirtzis:2020}. Unlike information technology (IT) systems, where vulnerabilities can lead to loss of information or privacy, vulnerabilities in the highly integrated information processing and physical control technology intrinsic to CPS could have public health and safety consequences~\cite{ars:hospital,technologyreview:triton,serpanos:2019}.

Perimeter-based security approaches (e.g., firewalls and encrypted communication channels) show some success in protecting CPS. However, a purely perimeter-based defense is asymmetric because attackers have the advantage of choosing the point of attack and access to many new kinds of vulnerabilities. In the context of CPS, perimeter security tends to be agnostic to the systems’ purpose, its required service or mission, and the functional behaviors of its cyber-physical aspects. Securing individual components is important in some contexts, but CPS are vulnerable to compromises in the interactions between components, even in the absence of what would traditionally be viewed as an individual component attack~\cite{amin:2013}. They are also vulnerable to supply chain and insider attacks~\cite{jones:2012}. 

Instead of attempting to react to, or predict, adversaries' specific capabilities, attack resilience is the ability of a system or domain to withstand attacks or failures, and in such events, to reestablish itself quickly~\cite{amin:2013,linkov:2013}. The goal of resilience (particularly for CPS) is to proactively assure the safety of the system by maintaining state awareness and physical system control. By first focusing on the safety of CPS, engineers and analysts can bound or focus the problem in ways that are challenging for pure IT systems.

There are many approaches to attack resilience, and it is currently an open problem and an active field~\cite{jones:2012,pajic:2014,nap:2019}. While many of these techniques are successful, each has a particular implementation and associated costs, as well as operational costs that involve both financial and performance tradeoffs. For a given solution, system designers and owners must understand the tradeoffs between costs, performance degradation, complexity, and improvements in resilience. To make matters worse, there is a combinatorial number of different solutions when one considers all of the possible solutions for a given function along with the collections of functions that are found in a CPS. The design of a CPS engenders a potentially intractable decision problem.

The question then becomes: what set of resilience-based solutions can be used, where in the CPS should these solutions be deployed, and in what combination? 

We argue that a systematic, tractable, and rigorous method is needed to support decision-making for implementing CPS resilience solutions. Designers of CPS must be able to manage the complexity of the decisions themselves and understand, and balance, the benefits and costs of resilience solutions. We have developed a framework called {\em Mission Aware} cybersecurity, which aims to manage complexity through general systems theory, framing CPS cybersecurity as a safety control problem. Mission aware supports decision-making through the use of three fundamental concepts: (1) CPS modeling based on systems theory and top-down hazard analysis; (2) automated vulnerability assessment via mining of attack databases; and (3) reusable design patterns, many of which exist in the literature and some of which have been developed by the authors. To explain and demonstrate these concepts, we develop an example based on an application to an unmanned aerial vehicle (UAV) performing a tactical reconnaissance mission.

\section{Managing Complexity through Abstraction}
Rather than beginning with tactical questions of how to protect a system against attacks, a strategic approach begins with questions about what essential services and functions must be secured against disruptions and how this disruption can lead to unacceptable loss. The specific implementation details will be used later to reason more thoroughly about only a subset of all the possible vulnerabilities; that is, only those combinations that can lead to specific undesirable outcomes. We argue that any resilience approach should move “top-down”, from general to specific, from abstract to concrete, from system-level goals and hazards to component-level behaviors and their interactions.

One of the powerful ways to manage complexity is by using hierarchical abstraction and refinement. By starting at a high level of abstraction with a small list of hazards or goals and simple models, and then refining that list and its associated models with more detail at each step, the stakeholders can be more confident about completeness and consistency of the analysis. This is because each of the longer lists of causes (refined hazards or causes) and more complex models (refined behaviors, analysis, and simulations) can be traced to the small starting list and models. With this approach, high-fidelity modeling, analysis and simulation is needed only on a subset of the CPS to provide assurance of correct behavior during deployment. By beginning with unacceptable or undesirable outcomes at the top (and not all possible outcomes), this approach reduces the total state space that one might need to explore at the lowest levels of abstraction.

Using the abstraction techniques described above, we can leverage historical vulnerability and weakness databases more effectively by basing search parameters on strategically relevant parts of the system. Historical vulnerabilities associated with those relevant system components help identify the potential threats to the system, and consequently, motivate the choice of particular resilience strategies to use in response to those threats.  

\subsection{UAV Mission Use Case}
We use as an example a UAV within a tactical reconnaissance mission that requires the UAV to produce data of the terrain, human activity, and aerial traffic within a particular area of interest. This mission involves identifying and localizing possible uncontrolled fires. Consequently, there is pressing need that the vehicle, the sensors that collect the data, and the data all maintain an acceptable level of performance in spite of potential adversarial actions. This mission is complex in that it involves a diverse set of components and technologies that are subject to a variety of potential threats. At the same time, UAV-based reconnaissance is a familiar scenario in many domains.  

\section{Systems and Graph Theory for Safety and Security}
{\em Mission Aware} involves an early system engineering process that identifies a high-level set of system objectives and unacceptable losses that represent system owners, operators, and other stakeholders~\cite{carter:2018,carter:2019}. Assuming that one has a high-level, comprehensive set of unacceptable outcomes, {\em Mission Aware} then involves constructing a model of the system from a control perspective based on the systems-theoretic accident model and process (STAMP)~\cite{young:2014}. 
Specifically, we identify the controllers, the actions available to them, and how those actions potentially lead to mission losses. 

STAMP is an accident causality model that captures accident causal factors including organizational structures, human error, design and requirements flaws, and hazardous interactions among non-failed components~\cite{leveson:2016}. In STAMP, system safety is reformulated as a system control problem rather than a component reliability problem: accidents occur when component failures, external disturbances, and/or potentially unsafe interactions among system components are not handled adequately or controlled. The safety controls intended to prevent such accidents are embodied in a hierarchical safety control structure, whereby commands or control actions are issued from higher levels and feedback is provided from lower levels. 

There is an important difference between STAMP and traditional hazard analysis techniques such as FMEA and FTA. The latter primarily focus on system failure (e.g. system reliability) as a function of individual component failures in the system, which is quantifiable via physical failure rates. Other potential causal factors, such as complex software errors and unsafe component interactions, are often not thoroughly considered. There is less technical agreement on quantifying security probabilities, where the origin of failure is from an adversarial act, a design vulnerability, or a misinterpretation of security requirements. For this reason, our approach stresses systematic methods to aid in the design and selection of resilience measures that are agnostic to the underlying probability of a successful attack.

\begin{figure*}[htb]
    \centering
    \includegraphics[width=\linewidth]{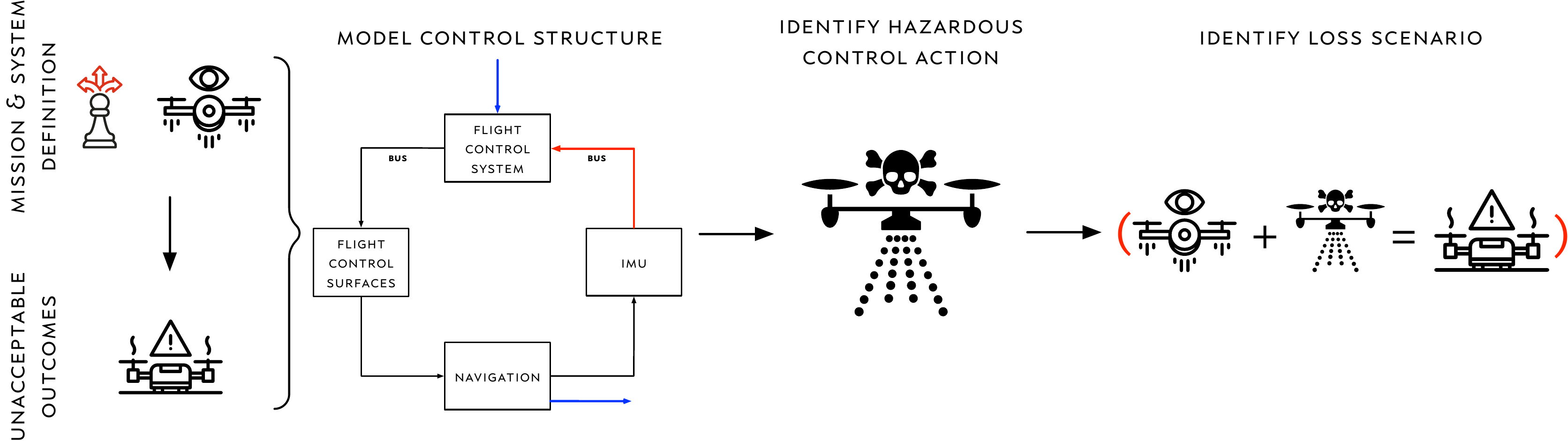}
    \caption{The STAMP-based modeling and analysis begins with defining the system and the mission it performs. Then, identifying unacceptable outcomes, the system’s control structure, and hazardous control actions help define the conditions that can lead to loss scenarios. If the conditions that lead to losses are identified, then we can implement safeguards in the form of resiliency or other solutions to prevent those conditions from occurring.}
    \label{fig:mission-aware}
\end{figure*}

The {\em Mission Aware} framework (Figure \ref{fig:mission-aware}) systematically encodes from the mission level all the way down to the hardware and component level: 
\begin{enumerate}
    \item the unacceptable outcomes of the mission, 
    \item the hazardous states that can lead to those outcomes, 
    \item the control actions that could lead to hazardous states, and circumstances under which those actions can create hazardous states, and
    \item the combinations of causes that can lead to hazardous control actions.
\end{enumerate}

\begin{figure*}[htb]
    \centering
    \includegraphics[width=.9\linewidth]{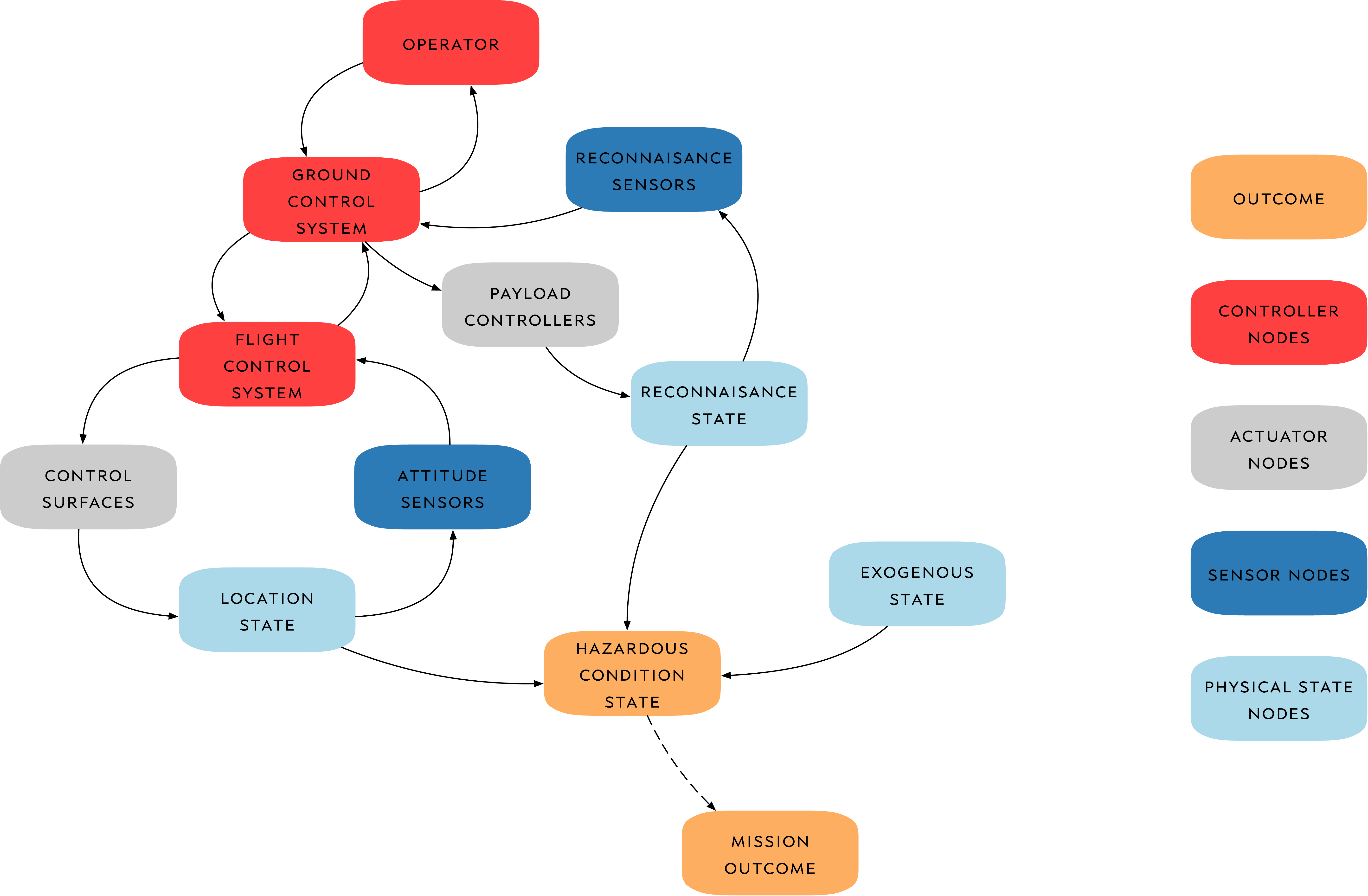}
    \caption{This is a portion of the specification graph for the reconnaissance UAV mission. The S-graph represents the system’s functional control structure in combination with the physical states that determine the presence or absence of undesirable outcomes. By using multiple ``types'' of nodes in the graph, we can combine behavior, consequences, and control structure in the same model.}
    \label{fig:s-graph}
\end{figure*}

This process allows for full top-to-bottom and bottom-up traceability, which supports evaluation of the cascading effects of specific changes to hardware, software, the order of operations, or other classes of behaviors on the potential outcome of a mission. This information is then used to identify vulnerable areas appropriate for resiliency or other security solutions. The pieces of information collected in the STAMP-based analysis are encoded into the system models (Figure \ref{fig:s-graph}), which are then used for further analysis and updated iteratively~\cite{carter:2018,carter:2019}. 

The specification graph (S-graph) combines diverse types of ``states'', or nodes, to represent the system operating in its mission environment. Valid decision behavior of the operator, hazardous conditions, and mission outcome nodes are encoded as truth tables, which perform the standard Boolean algebra on critical combinations of states in the system. Physical state nodes represent the set of variables that influence (or are influenced by) mission outcomes, while actuator and sensor nodes represent the system’s ability to manipulate and measure those physical states, respectively. The set of control actions or transition conditions between nodes are represented by edges between controllers or between a controller and its actuator(s). By using this graphical representation, we capture the traceability from the STAMP-based analysis between mission outcomes and individual component interactions, behaviors, or vulnerabilities.

STAMP can generate many hazardous scenarios~\cite{leveson:2012,leveson:2016}; we pick one simple example to illustrate traceability in the S-graph and how this traceability then provides focus for threat modeling and design.

\begin{figure*}[htb]
    \centering
    \includegraphics[width=.8\linewidth]{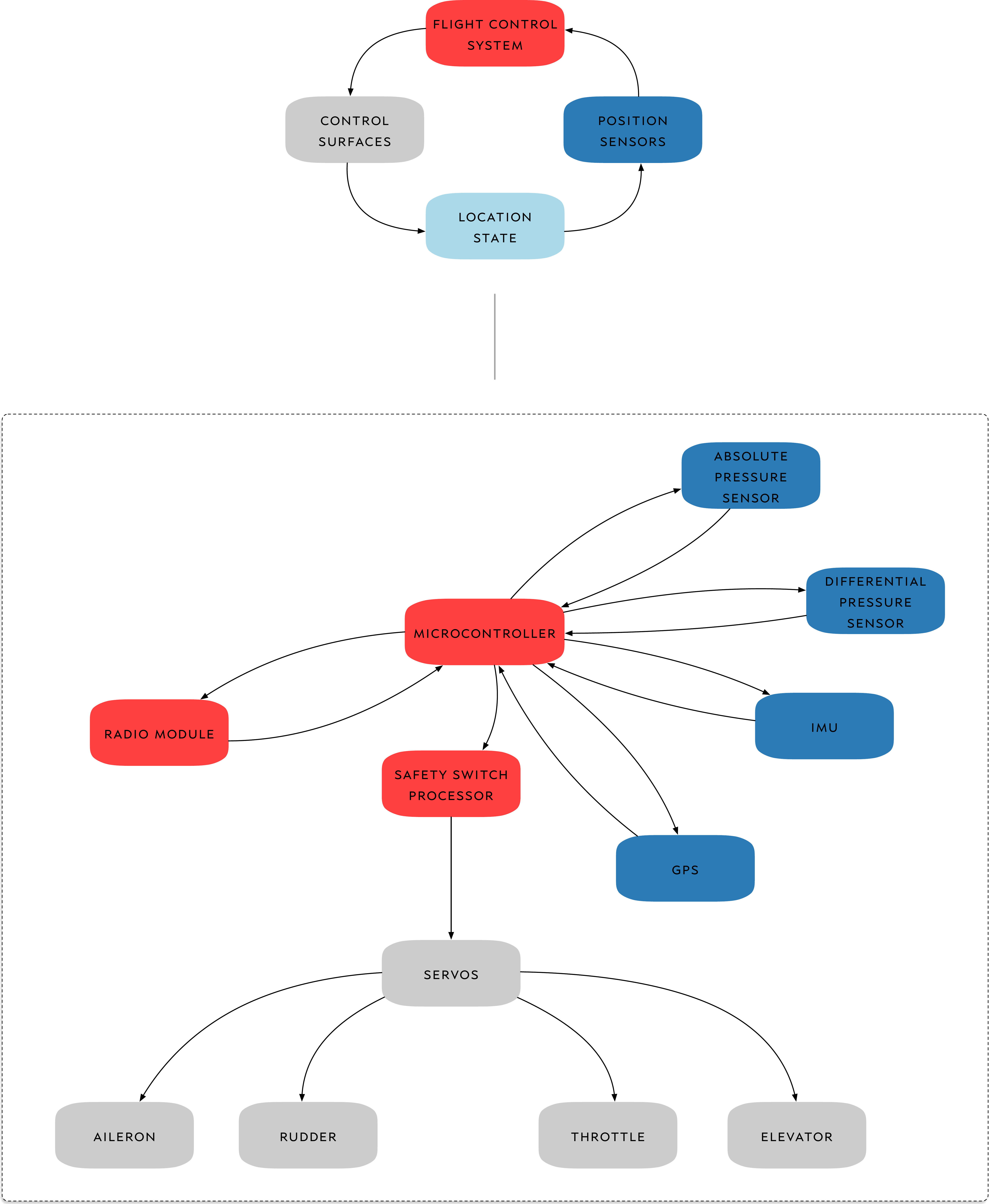}
    \caption{The S-graph is an initial formulation of subsystems that are critical to the mission. To develop a concrete threat model these critical subsystems must be further decomposed to a system architecture, which is one of many possible designs. The reason for the necessity of the decomposition is twofold: (1) to have a system model at the same level of detail necessary to match to attack vectors and (2) to produce metrics such as the system’s attack surface and exploit chains.}
    \label{fig:threat}
\end{figure*}

For the UAV example, mission stakeholders defined the mission, its goals, unacceptable outcomes, and other relevant operational insights. %
Let us assume that the most pressing loss is the loss of trustworthiness of the reconnaissance information chain based on the mission description. 
This scenario can be  illustrated as a situation where the fire services believe there is no fire near Waypoint A, when the UAV had imaged an area far from point A. The aircraft first navigates to an unrequested region and then captures information about that area by activating its imaging payload. This scenario involves otherwise ``correct'' control actions (the flight control system maintains stable control and the payload activates), but the UAV images the wrong area and sends this information to the operator. Incorrect mission decisions could result by the failure of the UAV to image fires in a vulnerable area. 

This example illustrates the notion that STAMP and the S-Graph can capture coupling and interdependencies between multiple functions and controllers within a system. Among many possible lower-level causes of this scenario, the navigation function of the UAV becomes critical, illustrating the coupling between the navigation components, the active control of the aircraft control surfaces, and the timing of when the imagery payload is activated. %

Here we map this informal example scenario to respective nodes and edges in the S-Graph (figure \ref{fig:s-graph}):
\begin{itemize}
    \item Mission loss (orange node): in appropriate allocation of suppression resources by fire service.
    \item Hazard (orange node): a combination of (incorrect) latitude-longitude values + activation of payload.
    \item Unsafe control action (grey nodes): 1. payload activated out-of-sequence with respect to 2. manipulation of control surfaces, leading to hazard above.
    \item Causal factors (attitude sensor node): inaccurate and/or delayed UAV location information.
    \item Resulting focus of threat modeling (next section): GPS and other navigation equipment/software.
\end{itemize}

At this point in the methodology, the model is agnostic to the initial cause of the scenario but can further be augmented with a concrete threat model. In turn, the mission-level requirements and threat model could better inform resilience or hardening defenses.

\subsection{Threat Modeling}

The S-graph -- as produced by the STAMP analysis -- results in finding and annotating the mission-critical subsystems in the initial model. From this analysis a threat model naturally emerges in the sense that analysts and designers can use the S-graph model to produce a list of the subsystems that, if exploited, could cause mission degradation. At this stage, however, the details present in the S-graph model is not at the right abstraction to further inspect if and how these subsystems could be exploited. Furthermore, a single list of elements is insufficient to produce metrics that can augment and inform a threat model such as attack surfaces or exploit chains. 

For these reasons, the S-graph needs to be modified to include particular implementation choices of software, network, and hardware that a designer is considering at the earlier stages of design (Figure \ref{fig:threat}). This extra information that is added to the S-graph is used to map subsystem elements to attack vector databases. Attack vector databases contain attack patterns, weaknesses, and vulnerabilities. Databases, such as CAPEC and CWE, of this form have little in the form of quantitative information but a lot in textual descriptions of exploits and their solutions. Therefore, to properly map entries related to the system under examination, it is necessary to add specific details to the model on system implementation choices.  Such choices could, e.g., be the type of hardware platform, type of OS, type of network connectivity, private or public networks.  By adding extra keywords to the model, the S-graph is now able to associate with attack vector databases like CAPEC and CWE~\cite{bakirtzis:2019}.

This additional information in the S-graph assists to semi-automate the process of finding possible exploits as well as constructing the attack surface by finding attacks for the entry point of a given subsystem. Tools and visualization methods using natural language processing can be used to aid in this process~\cite{adams:2018,bakirtzis:2018b}. As the system is refined, implementation choices with associated details emerge. It is at this phase in the lifecycle development where potential exploits or attack surfaces for the system architecture are captured.

A critical subsystem might not be immediately accessible by attackers, but a series of attacks starting from the network accessible elements of the system and continuing by exploiting a series of connected subsystems can manage to reach and exploit a critical subsystem. For example, the microcontroller of a UAV might not be immediately attackable but there exists a possible series of attacks that start with the radio modules that does lead to the microcontroller being exploited. Another more direct example is the exploitation of the GPS that is used to provide positioning to the flight controller. The mapping between the system model and attack vector databases in this case produces CVE-2016-3801, which describes a vulnerability in the Mediatek GPS drivers for Android One devices. This specific vulnerability may not directly affect a UAV system; however, it is an instantiation of CWE-264, Permissions, Privileges, and Access Control, which is not immediately produced by the information added by system designers but can be (automatically) identified because of the hierarchical nature of attack vector data. This broader class of weaknesses gives system designers, integrators, and operators a basis from which to be aware of the risk associated with specific implementation choices.

This threat modeling analysis can be conducted at various times in the system lifecycle as different levels of system and component details emerge from preliminary design. The benefit of this type of threat modeling is that it is systematic and traceable. Analysts and designers can augment system requirements, modify the system architecture, or implement secure design principles to reflect the findings of threat posture analysis. The stopping point is when a complete threat model is drafted with respect to the proposed system architecture containing attack vectors that map to system elements. The attack surface resulting from the analysis informs analysts where attacks can enter the system and outlines exploit chains. By producing a complete threat model only based on system models, it is possible to incorporate security analyses earlier in the lifecycle and, therefore, inform security design decisions at a stage where  design changes have much less of a cost impact. Such actions could be security focused, e.g. opting for more secure hardware and software platforms,  or resilience focused where redundancy, component diversity and recovery principles are employed to secure the architecture.

\section{Design Patterns for Resilience}

As suggested in 2018 NAE workshop on \emph{cyber resilience}, system level patterns for implementing resilience are needed~\cite{nap:2019}. In the {\em Mission Aware} approach, system enhancements take the form of reusable design patterns (either physical, software, or procedural) that intend to increase resilience. The set of all possible design patterns is large for even a simple system, and it is possible that several of the possible design patterns would have little effect on the resilience of the system. Nonetheless, one can describe most resilience design patterns in terms of several well-known principles that follow from system self-healing and self-protecting attributes from autonomic computing: (1) diversification, (2) redundancy, (3) randomization, and (4) system policy adaptation~\cite{dobson:2010}. Design patterns using these principles adapt the system to increase the effort required to successfully attack and compromise the system. It should be noted that adaptation might create unintentional effects that degrade system performance, an observation that reinforces the need to explicitly model the requirements or objectives of the system. 

Redundancy is a common design paradigm where multiple critical components are used to perform the same function so that if one fails another can take its place.  For a resilience system, diverse redundancy requires that two or more components that perform the same function are diverse with regards to a common attack pattern. It is imperative that the components be diverse because, even with redundancy, a common source of failure (e.g. a successful attack) could cause all the redundant components to fail.  This design pattern mitigates the ability for a single successful attack to compromise all redundant components.    

A redundant configuration of the flight control system in the UAV example mitigates the risk of vehicle loss due to the failure of a single controller. However, if a supply chain attack successfully embeds a Trojan horse onto the controller, then redundancy is not sufficient. While the system is resilient to natural failures, this single attack can compromise all controllers. The diverse redundancy solution is to procure each controller from a different supplier, thus mitigating the risk of an insider supply chain attack.

Similar to diverse redundancy, {\em verifiable voting}~\cite{jones:2012} uses redundant components to confirm the output of a system. Each component is a voting mechanism implemented in software or hardware and should be simple enough to be secured. If the voting mechanisms do not agree on an output, it is likely that an attack, or some other fault, has occurred. Verifiable voting allows for the system to remain in use when under attack. If it can be confirmed that only one voting mechanism is compromised, the results from the other voting mechanism can still be used. For example, the UAV’s primary source of information may be a standard camera. The data from the camera is transmitted to a media server and then relayed via a wireless signal to interested parties. The media server is vulnerable to an insider attack. A verifiable voting solution to this vulnerability is to install a secondary camera payload with lesser but acceptable performance and a second media server in a separate location. The redundant system monitors the same information but mitigates the risk of attack by supplying redundant information to the interested parties.

{\em Physical configuration hopping}~\cite{jones:2012} is another design pattern for resilience derived from the idea of redundant systems. As in diverse redundancy and verifiable voting, several redundant components are given the same objective in a system. However, when implementing physical configuration hopping, control and execution is randomly moved between the redundant components. Physical configuration hopping can be combined with diverse redundancy to further mitigate the risk of an insider supply chain attack. In the UAV, physical configuration hopping can be implemented by randomly hopping control of location monitoring between the redundant flight controllers. Virtual configuration hopping is similar to physical configuration hopping; however, hops occur between virtual components instead of physical components.

\section{Evaluating Risk and Tradeoffs}

As with any other design choice, resilience requires the ability to characterize and evaluate the tradeoff between potential gain in resilience and the cost of a proposed design. The cost of a design pattern can take on many notions including the financial cost of acquiring and installing hardware, the cost of increased complexity, and the cost of operational degradation. While some of these costs can be measured and quantified, quantifying the gain in resilience is more difficult and still an open question. Benefits of implementing security or resilience solutions include eliminating or reducing the possibility of attack, or mitigating and containing the results of a successful attack. In particular, reasoning about resilience should be defined as a function of three variables or dimensions: (1) the severity of mission-level outcomes, (2) the complexity of the attack vectors needed to achieve the outcome, and (3) the cost and complexity of mitigating such attacks.

Consequence can only be determined (and ranked) by the owners and other stakeholders of the system, and these consequences have to be agnostic to implementation details, architectures, or threats. The key step is then providing a clear mapping between component vulnerabilities and system-level consequences, and vice-versa. We provide one approach to obtain this information and mapping in the {\em Mission Aware} section above utilizing STAMP concepts and the S-graph. This represents one of the dimensions necessary to reason about designing resilience into systems. 

The next dimension, attack complexity, requires analysis of possible adversaries and their available techniques. 
Of course there is a risk in not identifying all available techniques, which we mitigate by focusing on (a) resilience and not prevention and (b) on system-level consequence, not likelihood. Given that this dimension is based on difficult-to-quantify assessments of attacker profiles and the techniques available to them, attack complexity is (currently) developed from expert opinion based on historical evidence.
The last dimension, mitigability, is a function of the mission itself, its system architecture, and the design patterns available to handle a set of threats (i.e. the information derived from the ``Design Patterns'' section above). Currently, this dimension is based on expert opinion due to the number of disparate factors that contribute to this measure. A scoring method could be applied to compare different resilience solutions; however, it is likely that the appropriate weights and scoring functions vary from design to design.

Based on the criticality of maintaining location integrity in the UAV mission and the vulnerability assessment revealing the threat of attackers exploiting permission, privileges, and access control vulnerabilities in the UAV, a possible resilience strategy would be physical configuration hopping in the flight control system. This solution would significantly increase attack difficulty and workload for the adversary due to the changing attack surface. Additionally, because the mission is dependent on visual imagery being correctly linked to its location, diverse redundancy and verifiable voting are natural candidates. Implementation could be an entirely different type of GPS device or even a different class of navigation components (e.g. inertial navigation). This pattern does not necessarily prevent or mitigate attacks of the GPS but make mitigation of a successful GPS attack more effective; the system can revert to another mode of navigation if the attack is detected.

\section{Conclusions}
In summary, application of a pattern may increase the complexity of an attack required to yield an adverse outcome, and thus make the outcome less likely; or it may make the attack more mitigable and thus make the outcome less severe.  In some cases, a design solution may achieve both, or multiple design patterns can be used to address the same risk, as the example above illustrates. In any of these cases, it is also required to think about cost. In the case of multiple redundant GPS units, it might not be the extra component that is ``expensive'', although that may also be the case. Rather, one must design and assess the voting scheme, assuring that it is itself secure from attack. Our position is that the monitoring functions for these design patterns can and should be made as simple as possible, with the implication being that something simple is easier to secure (e.g., through formal verification or more complete testing).

Given that there may be multiple vulnerabilities in critical pathways, with multiple combinations of available design patterns, the system owners must make a tradeoff between their perceived risk of attack and its impact against the increased resilience the solutions offers. Of possible future research directions, developing metrics to quantify these tradeoffs may be some of the most critical work for advancing the {\em Mission Aware} methodology and the resilience of CPS as a field of study.

Critical CPS  are becoming much more common in daily life, and better ways of securing them are essential. The state of
practice for securing CPS is at a point where new system-oriented methods and tools are needed to adequately protect critical systems against advanced threats.


\bibliographystyle{IEEEtran}
\bibliography{computer}

\end{document}